\documentclass[a4paper,aps,onecolumn,showpacs,showkeys,nofootinbib,preprintnumbers,superscriptaddress,amsmath,amssymb,amsfonts]{revtex4}
\usepackage{graphicx}
\usepackage{bm,natbib}
\usepackage{amsmath}
\usepackage[english]{babel}
\usepackage[T1]{fontenc}
\usepackage{amssymb}
\usepackage{amsfonts}
\usepackage{epsfig}
\usepackage{colordvi}
\usepackage{psfrag}
\usepackage{color}
\usepackage{ mathrsfs }

\newcommand{\Lagr}{\mathcal{L}}
\newcommand{\G}{\mathcal{G}}

\usepackage{dcolumn}
\usepackage{multirow}
\usepackage{hyperref}
\usepackage{epstopdf}
\usepackage{bm}
\usepackage{times}

\hypersetup{
  colorlinks=true,        
  linkcolor=blue,         
  citecolor=magenta,      
}

\begin{document}

\title{Non-Local  Curvature and Gauss--Bonnet  Cosmologies by Noether Symmetries}

\author{Francesco Bajardi}
\email{bajardi@na.infn.it}
\affiliation{Department of Physics ``E. Pancini'', University of Naples ``Federico II'', Naples, Italy.}
\affiliation{INFN Sez. di Napoli, Compl. Univ. di Monte S. Angelo, Edificio G, Via Cinthia, I-80126, Naples, Italy.}

\author{Salvatore Capozziello}
\email{capozziello@na.infn.it}
\affiliation{Department of Physics ``E. Pancini'', University of Naples ``Federico II'', Naples, Italy.}
\affiliation{INFN Sez. di Napoli, Compl. Univ. di Monte S. Angelo, Edificio G, Via	Cinthia, I-80126, Naples, Italy.}
\affiliation{Scuola Superiore Meridionale, Largo San Marcellino 10, I-80138, Naples, Italy.}
\affiliation{Laboratory for Theoretical Cosmology, Tomsk State University of
Control Systems and Radioelectronics (TUSUR), 634050 Tomsk, Russia.}

\author{Daniele Vernieri}
\email{daniele.vernieri@unina.it}
\affiliation{Department of Physics ``E. Pancini'', University of Naples ``Federico II'', Naples, Italy.}
\affiliation{INFN Sez. di Napoli, Compl. Univ. di Monte S. Angelo, Edificio G, Via	Cinthia, I-80126, Naples, Italy.}

\date{\today}

\begin{abstract}
Non-local gravity cosmologies are considered under the standard of Noether Symmetry Approach. In particular, we focus on  non-local theories whose gravitational actions depend on curvature and  Gauss--Bonnet  scalar invariants.
Specific functional forms of the related point-like Lagrangians are selected by  Noether symmetries and we solve the corresponding field equations finding out exact cosmological solutions.  
\end{abstract}



\maketitle

\section{Introduction}

Soon after the formulation of General Relativity (GR) became  evident, despite several attempts, the quantization procedure could not be applied in the same way as in other field theories. Indeed, other fundamental interactions are now perfectly treatable under the standard of \emph{Gauge Theories}, although initially they presented some problems as well. From the path integral quantization procedure, it turned out that they diverge at ultraviolet (UV) scales; this (apparently) big issue was solved by means of the \emph{Renormalization} procedure, thanks to which it is now possible to eliminate UV divergences and fully quantize the theory~\cite{Weinberg:1995mt, Peskin:1995ev}. By introducing in the theory suitable cutoffs, it  happens that the $n$-point amplitude (cutoff dependent) converges even by taking the limit of infinite cutoff and re-summing at any level of the perturbation theory. In this case, the theory is said to be \emph{renormalizable}. However, it may be complicated to evaluate the amplitude at any $n$-loop level, as well as solving the renormalization group flow equation. In order to test the renormalizability of the theory, the mass dimension of the coupling constant plays a crucial role, allowing to check whether the theory is power-counting renormalizable (for non-negative coupling constant) or non-renormalizable. In this latter case, the method is not able to give precise information and, in order to check the renormalizability, the $n$-loop calculations on the $n$-point amplitude is needed.

In any case, if Quantum Field Theory holds even beyond the Planck scale, we have to adapt its formalism to the gravitational interaction, though, so far, GR cannot be treated as a renormalizable quantum field theory. This is partially due to the coupling constant mass dimension, which turns out to be negative after expanding the action around the Gaussian fixed point~\cite{Shomer:2007vq}. A possible way out of these shortcomings may be provided by modified theories of gravity which extend the gravitational action of GR and seem to solve issues at UV and infra-red (IR) scales.
In this context, non-local terms can be added to GR in view of an effective theory consistent with quantum dynamics.

 There are several GR extensions which, for example, could fix some  $\Lambda$CDM issues in cosmology (see \emph{e.g.}~\cite{delaCruz-Dombriz:2018nvt, Caruana:2020szx, Bhattacharjee:2020eec, Bajardi:2020fxh}, for other alternative theories of gravity), as the $f(R)$ theories of gravity~\cite{Nojiri:2017qvx, Capozziello:2009nq, Capozziello:2011et}, that introduce in the action  general functions of the Ricci scalar and do not need to invoke any Dark Energy  and Dark Matter fluids to explain the current universe expansion and large scale structure~\cite{Capozziello:2012ie, Hu:2007nk}. Among these theories,  the Starobinsky model~\cite{Starobinsky:2007hu}, based on Ricci scalar corrections,  seems an extremely reliable candidate to describe inflationary epoch. 
 
 Furthermore, several papers take into account actions containing other second-order curvature invariants~\cite{Calmet:2017rxl, Cognola:2012jz} like $f(R, R^{\mu \nu} R_{\mu \nu}, R^{\mu \nu p \sigma} R_{\mu \nu p \sigma})$. A particular important case of this latter function is given by the Gauss--Bonnet (GB) topological invariant. It consists in a  combination of second order curvature invariants and, specifically, is defined as 
 \begin{equation}
 \G:= R^2 - 4R^{\mu \nu} R_{\mu \nu}+ R^{\mu \nu p \sigma} R_{\mu \nu p \sigma}\,.
 \end{equation}
 The importance of the scalar $\G$ comes from the  \emph{Generalized Gauss--Bonnet Theorem} which states that the integral of the Gauss curvature over a given Manifold is equal to the Euler Characteristic~\cite{Nakahara:2003nw}. In four dimensions (or less) indeed, the GB scalar is nothing but a topological surface term and the action $S = \int \sqrt{-g} \; \G \; d^4x$ is everywhere trivial. Nevertheless, a function of the GB term turns out to be able to provide physical predictions even in $(3+1)$-dimensions. Usually, in order to get the Einstein--Hilbert action as a particular limit, the action $S =\int \sqrt{-g} \;\left\{R+ f(\G)\right\} \; d^4x$ is considered~\cite{Capozziello:2014ioa, Paul:1990jg, Paolella, Terrucha:2019jpm}. In this way, as the scalar curvature is predominant at local scales, the $f(\G)$ theory might  occur at  cosmological scales. 
 As we shall  discuss below, introducing a topological term in the theory allows to reduce the dynamics and to obtain reliable cosmologies.

 In general, non-local gravity is a class of theories aimed at describing gravity at low and high energy scales~\cite{Mashhoon:2011mb, Belgacem:2017cqo}. A general non-local action is supposed to contain functions of the inverse of  D'Alembert operator, leading to a positive mass dimension of the coupling constant and making the effective theory renormalizable~\cite{Modesto:2017hzl, Tomboulis:2015esa}. In Ref.~\cite{Nojiri:2007uq} for instance, a non-local theory containing only the scalar curvature is treated, and the authors also show that the theory is capable of providing a model for the early-time inflationary universe without introducing any further scalar field. Other cosmological implications of non-local theories of gravity are also studied in Ref.~\cite{Elizalde:2018qbm}, where the authors find both cosmological de Sitter and power-law expansion of the universe, after focusing on some specific actions. Specifically, in this latter case, a  non-local function of the GB topological invariant and of the Ricci scalar is studied, with the aim to constrain the theory by cosmological observations. 
 
 In this paper,  we are going to consider   cosmologies coming from non-local actions containing Ricci and   GB scalar invariants. The aim is to fix the form of non-local gravity Lagrangians by means of the Noether Symmetry Approach. As we shall see, the existence of such symmetries allows to reduce dynamics and to obtain exact cosmological solutions. 
 
 The layout of the paper is the following. After a brief summary on non-local theories of gravity in Sec.~\ref{sectnonlocal},
 we apply the Noether Symmetry Approach to non-local models containing  the Ricci curvature scalar in Sec.~\ref{non-local Einstein Gravity}. The selected Lagrangians are power-law in the Ricci scalar and exponential in the non-local terms. This fact is particularly interesting because, as demonstrated in~\cite{Modesto:2017hzl,Tomboulis:2015esa, Bahamonde:2017sdo}, this kind of term allows the regularization and renormalization of the theory. However, in this paper they are not introduced \emph{by hand} but selected by symmetries. Finally exact cosmological solutions are  derived. Sec.~\ref{Purenon}  deals with the application of the method to models containing only non-local terms of the topological invariant $\G$. We show that  GR is recovered in the special case $f(\G)\sim \sqrt{\G}$ (see also~\cite{Bajardi:2020osh}). Furthermore, it is shown that considering a Einstein--Hilbert action corrected with non-local terms of $\G$ is just a subcase of the more general theory $f(\G,\Box^{-1}\G)$.  Conclusions are drawn in Sec.~\ref{Conclusions}. The Noether Symmetry Approach is sketched in Appendix \ref{Noether Symmetry Approach}.

\section{Non-local Theories of Gravity} \label{sectnonlocal}

Let us review now the basic aspects of non-local theories of gravity. First of all, they can be divided into  two main categories: Infinite Derivative Theories of Gravity (IDGs) and Integral Kernel Theories of Gravity (IKGs). The former are used to be exponential functions of the D'Alembert operator \footnote{In general, they can  be also trascendental functions of differential operators.} and to overcome UV shortcomings by means of a short-range non-locality. The latter mainly involves the inverse of the D'Alembert operator \footnote{They can  involve integral kernels of differential operators} $\Box^{-1}$  and, by means of long-range non-locality, it is capable of fixing, in principle,  the IR problems of GR. The models treated in this paper involve  functions of the operator $\Box^{-1}$, which will be applied to the Ricci scalar $R$ and the GB invariant $\G$. For this reason, we outline only the properties of IKGs.

In general, the local corrections come from an expansion around the value  $s=0$ of a Schwinger proper time, so that they are valid for small times only, providing UV corrections. On the other hand, IR corrections are represented by the expansion around $s\to\infty$, where the proper time integration becomes divergent. This problem can be solved by considering a non-perturbative approach to calculate the Schwinger proper time integral which allows to capture both the effects of local UV contributions ($s=0$) and of non-local IR effects ($s\to\infty$).

The corresponding quantum effective action in curved spacetime reads~\cite{Barvinsky:2014lja}:
\begin{equation}
W_0=-\int\!d^4x \sqrt{-g}\,\Bigl[V(x)+V(x)(\square-V)^{-1}V(x)\Bigr]+\frac{1}{6}\,\Sigma \,,
\end{equation}
where $V(x)$ is the potential and $\Sigma$ a surface term defined as
\begin{equation}
\begin{split}
\Sigma=\int\!d^4x \sqrt{-g}\,\Bigl\{R&-R_{\mu\nu}\,\square^{-1}G^{\mu\nu} %
                                           2^{-1} R\,\bigl(\square^{-1}R^{\mu\nu}\bigr) \square^{-1} R_{\mu\nu} \\
             &-R^{\mu\nu}\bigl(\square^{-1} R_{\mu\nu}\bigr) \square^{-1}R %
                \bigl(\square^{-1} R^{\alpha\beta}\bigr) \bigl(\nabla_{\!\alpha}\,\square^{-1}R\bigr) \nabla_{\!\beta}\,\square^{-1}R \\
             &-2\,\bigl(\nabla^{\mu}\,\square^{-1} R^{\nu\alpha}\bigr) \bigl(\nabla_{\!\nu}\,\square^{-1} R_{\mu\alpha}\bigr) \square^{-1}R \\
             &-2\,\bigl(\square^{-1} R^{\mu\nu}\bigr) \bigl(\nabla_{\!\mu}\,\square^{-1}R^{\alpha\beta}\bigr) %
                 \nabla_{\!\nu}\,\square^{-1} R_{\alpha\beta}+O\bigl[R_{\mu \nu}^{\,\,4}\bigr]\Bigr\} \,.
\end{split}
\end{equation}
The integral operator $\square^{-1}$ is the responsible for quantum corrections to GR. A simple action containing such an operator was proposed by Deser and Woodard in Ref.~\cite{Deser:2007jk}, where they presented a non-local modified effective theory of gravity capable of explaining the current late-time cosmic acceleration as a mechanism driven by the integral kernel of some differential operator; the corresponding action reads:
\begin{equation}
\label{NL:8}
\mathcal{S}=\int\!d^4x \sqrt{-g}\,\,\Bigl[\frac{R}{2 \kappa}+F\Bigl(\square^{-1}R\Bigr)\Bigr] +\mathcal{S}^{(m)}\,,
\end{equation}
where $\kappa=8\pi G_N$, and $F\Bigl(\square^{-1}R\Bigr)$ is an arbitrary function of $\square^{-1}R$. The field equations associated to the effective theory~\eqref{NL:8} are
\begin{equation}
G_{\mu\nu}+\kappa \Delta G_{\mu\nu}=\kappa T_{\mu\nu}^{\,(m)} \,,
\end{equation}
where
\begin{equation}
\begin{split}
\Delta G_{\mu\nu}=\,&\Bigl(G_{\mu\nu}+g_{\mu\nu}\,\square-\nabla_{\mu}\nabla_{\nu}\Bigr)%
                                \biggl\{F+\square^{-1}\Bigl[R\,F_{\Box^{-1}R}\Bigr]\biggr\} \\
                             &+\biggl[\delta_{\mu}^{\,\,(\rho}\,\delta_{\nu}^{\,\,\,\sigma)}-\frac{1}{2}\,g_{\mu\nu}g^{\rho\sigma}\biggr]%
                                \partial_{\rho}\Bigl(\square^{-1}R\Bigr)\,\partial_{\sigma}\biggl(\square^{-1}\Bigl[R\,F_{\Box^{-1}R}\Bigr]\biggr)                  \,,
\end{split}
\end{equation}
with the definitions $F\equiv F\Bigl(\square^{-1}R\Bigr)$ and $\displaystyle F_{\Box^{-1}R} \equiv\frac{\partial F}{\partial\bigl(\square^{-1}R\bigr)}$ . It is straightforward to show that the intrinsic nature of  operator $\Box^{-1}$ is able to predict naturally  the late-time cosmological expansion of the universe. In this regards, considering a power-law form of the scale factor given by the standard cosmological models
\begin{equation}
\label{NL:19}
a(t)\sim t^q \quad\implies\quad R(t)\sim6q(1-2q)\,t^{-2}\,,
\end{equation}
it is possible to approximately evaluate the quantity $\bigl(\square^{-1}R\bigr)(t_0)$ at the present time. Indicating with $t_0\sim10^{10} \, y$ the current time and with $t_{eq}\sim10^5 \, y$ the time when the CMB radiation originated, the non-local causal effects acting during the interval $[t_{eq},t_0]$ are: 
\begin{equation}
\begin{split}
\bigl(\square^{-1}R\bigr)(t_0)&=\int_{t_{eq}}^{t_0}\!\!dt'\frac{1}{a^3(t')}\int_{t_{eq}}^{t'}\!\!dt''a^3(t'')R(t'')=\\
                                            &=\frac{6q(2q-1)}{(3q-1)}\Biggl\{\log\biggl(\frac{t_0}{t_{eq}}\biggr)-\frac{1}{3q-1}+\frac{1}{3q-1}%
                                                 \biggl(\frac{t_{eq}}{t_0}\biggr)^{3q-1}\Biggr\}.
\end{split}
\end{equation}
Taking into account a standard matter dominated universe with $q=2/3$,  we have
\begin{equation}
\bigl(\square^{-1}R\bigr)(t_0)\bigl|_{q=\frac{2}{3}}\,\,\sim 14.0\,\,.
\end{equation}

The above result suggests that the non-local term leads to the order required by the current cosmic acceleration and avoids the fine tuning of parameters. Furthermore, these corrections  occur only at late-times, since during the radiation dominated era the non-local effects are null while, after the onset of matter dominance, the logarithmic dependence make them negligible.

\section{Noether Symmetries in Non-Local Curvature Cosmology} \label{non-local Einstein Gravity}

With the above considerations in mind, let us see if the form of the non-local action containing the operator $\Box^{-1}$ applied to the Ricci scalar $R$ can be selected by the Noether Symmetry Approach (see~\cite{Capozziello:1996bi, Vakili:2008ea, Sadjadi:2012xa, Sanyal:2003tv, Bahamonde:2018ibz, Capozziello:2013qha, Capozziello:2012hm, Tsamparlis:2011wf, Tsamparlis:2011zz, Dialektopoulos:2018qoe, Bajardi:2020xfj, Capozziello:2012iea, Capozziello:2015hra, Bajardi:2019zzs, Urban:2020lfk} for details and applications). An extended discussion including this case is treated in Ref.~\cite{Bahamonde:2017sdo}.
A straightforward generalization of the action \eqref{NL:8} is:
\begin{equation}
S = \int \sqrt{-g} f(R,\Box^{-1} h(R)) \; d^4x \,.
\label{initial action R}
\end{equation}
Starting from this, we want to  derive the  cosmological point-like Lagrangian and then search for   Noether symmetries according to the lines sketched in App. \ref{Noether Symmetry Approach}. The first issue  is to define a suitable {\it  localization} of the non-local field. We can define:
\begin{equation}
\label{local}
\Box^{-1} h(R) := \phi \;\;\;\;\;\;\; \to \;\;\;\;\;\;\; h(R) = \Box\phi\,,
\end{equation}
where $\phi$ is an auxiliary scalar field. Let us now  focus on a  spatially-flat Friedmann--Lema\^itre--Robertson--Walker (FLRW) cosmological background, with metric $ds^2 = dt^2 - a(t)^2 d\textbf{x}^2$, where $a(t)$ is the scale factor of the universe depending on the cosmic time $t$. In this perspective, we have also that the scalar field depends only on $t$, that is $\phi=\phi(t)$. The Ricci scalar can be expressed as
\begin{equation}
 R= -6 \left( \frac{\ddot{a}}{a} + \frac{\dot{a}^2}{a^2} \right)\,,
 \end{equation}
 and  can be directly introduced into  action \eqref{initial action R}.  Considering also the localization \eqref{local}, we can define  the further  scalar field $\epsilon(t)$ such that  action \eqref{initial action R} can be written as
\begin{equation}
S= \int \left\{a^3 f(R, \phi) + a^3 \epsilon(t)\left[\ddot{\phi} + 3 \frac{\dot{a}}{a} \dot{\phi} - h(R)\right] - \lambda \left[R + 6 \left( \frac{\ddot{a}}{a} + \frac{\dot{a}^2}{a^2} \right)\right] \right\}dt\,.
\end{equation}
From the variation with respect to $R$, we  find the Lagrange multiplier $\lambda$, that is:
\begin{equation}
\frac{\delta S}{\delta R} = \int \left\{ a^3 f_R(R,\phi) - a^3 \epsilon(t) h_R(R) - \lambda\right\} dt = 0 \;\;\;\;\; \to \;\;\;\;\;\; \lambda =  a^3 \left[f_R(R,\phi) - \epsilon(t) h_R(R) \right]\,,
\end{equation}
where the subscript $R$ denotes the derivative with respect to the curvature scalar. 
The variation of the action with respect to the scalar field $\epsilon(t)$ provides the Klein-Gordon equation $\Box \phi = h(R)$  as expected. Finally, by varying the action with respect to the scalar field $\phi$,  we get:
\begin{equation}
\Box \epsilon(t) = f_\phi(R,\phi)\,.
\end{equation}
The canonical point-like Lagrangian arising after an integration by parts  is then:
\begin{eqnarray}
\Lagr &=& a^3 \left( f(R,\phi) - R f_R (R,\phi) - \epsilon h(R) + \epsilon R h_R(R) \right) - a^3 \dot{\epsilon} \dot{\phi} - 6 a \epsilon \dot{a}^2 h_R(R) - 6 a^2 \dot{a} \dot{\epsilon} h_R(R) \nonumber
\\
&& - 6 a^2 \epsilon \dot{a} \dot{R} h_{RR}(R) + 6 a \dot{a}^2 f_R(R,\phi) + 6 a^2 \dot{a} \dot{\phi} f_{R \phi}(R,\phi) + 6 a^2 \dot{a} \dot{R} f_{RR}(R,\phi) \,.
\label{R Lagr}
\end{eqnarray}
The corresponding Euler--Lagrange equations and the energy condition are, respectively:
\begin{eqnarray}
\frac{d}{dt} \frac{\partial \Lagr}{\partial \dot{a}} = \frac{\partial \Lagr}{\partial a} \;\; \to \;\; &&  2 \dot{a}^2 \left(f_R(R,\phi)-\epsilon h_R(R) \right) - 4 a \left[ \ddot{a}\left( \epsilon h_R(R) - f_R(R,\phi)  \right) + \dot{a} \left(\dot{\epsilon} h_R(R) + \epsilon \dot{R} h_{RR}(R) - \dot{\phi} f_{R \phi}(R,\phi)  \right. \right. \nonumber  
  \\
  && \left. \left. - \dot{R} f_{RR}(R,\phi) \right)\right] + a^2 \left[- f(R,\phi) + \dot{\epsilon} \dot{\phi} - 2 h_R(R) \ddot{\epsilon} - 4 \dot{R} \dot{\epsilon} h_{RR} + \epsilon \left(h(R) - R h_R(R) - 2 \ddot{R} h_{RR}   \right. \right.   \nonumber
   \\
   && \left.  - 2 \dot{R}^2 h_{RRR}(R) \right) + R f_R(R,\phi) + 2 \ddot{\phi}f_{R \phi}(R,\phi) + 2 \dot{\phi}^2 f_{R \phi \phi} (R,\phi) + 2 \ddot{R} f_{RR}(R,\phi)   \nonumber
   \\
  &&\left. + 4 \dot{R} \dot{\phi} f_{RR \phi} (R,\phi) + 2 \dot{R}^2 f_{RR}(R,\phi)\right] = 0\,; \nonumber
\end{eqnarray}
\begin{eqnarray}
\frac{d}{dt} \frac{\partial \Lagr}{\partial \dot{\phi}} = \frac{\partial \Lagr}{\partial \phi} \;\; \to \;\; && \Box \epsilon = f_\phi(R, \phi)\,; \hspace{12.5cm} \nonumber
\end{eqnarray}
\begin{eqnarray}
\frac{d}{dt} \frac{\partial \Lagr}{\partial \dot{R}} = \frac{\partial \Lagr}{\partial R} \;\; \to \;\; && R= -6 \left(\frac{\ddot{a}}{a} + \frac{\dot{a}^2}{a^2} \right); \hspace{11.7cm} \nonumber
\end{eqnarray}
\begin{eqnarray}
\frac{d}{dt} \frac{\partial \Lagr}{\partial \dot{\epsilon}} = \frac{\partial \Lagr}{\partial\epsilon} \;\; \to \;\; && \Box \phi = h(R)\,; \hspace{12.9cm} \nonumber
\end{eqnarray}
\begin{eqnarray}
E_{\Lagr}=\dot{q}^i \frac{\partial \Lagr}{\partial \dot{q}^i} - \Lagr = 0 \;\; \to \;\; && 12 \dot{a}^3 \left(f_R(R,\phi)- \epsilon h_R(R) \right) - a^3 \left[f(R,\phi) - \epsilon \left(h(R) - R h_R(R) \right) \right]  \nonumber 
\\
&& +a^3 \dot{\epsilon} \dot{\phi} - a^3 \ddot{\epsilon} \dot{\phi} - a^3 \dot{\epsilon} \ddot{\phi} + a^3 R f_R(R,\phi) + 6 a \dot{a} \left[ \epsilon \left( -2 h_R(R) \ddot{a} + \dot{a} h_R(R)   \right. \right.  \nonumber
\\
&&\left. \left. - 4 \dot{a} \dot{R} h_{RR}(R) \right)+ 2 \ddot{a} f_R(R,\phi) - \dot{a} \left(6 \dot{\epsilon} h_R(R) + f_R(R,\phi) - 6 \dot{\phi} f_{R \phi}(R,\phi) \right. \right.  \nonumber
\\
&& \left. \left. - 4 \dot{R} f_{RR} (R,\phi) \right) \right] + 6 a^2 \ddot{a} \left(\dot{\phi} f_{R \phi}(R,\phi) - \dot{\epsilon} h_R(R) \right) + 6 a^2 \dot{a} \left[-\ddot{\epsilon} h_R(R)  \right. \nonumber
\\
&&+ \epsilon \dot{R} h_{RR}(R)  + \dot{\epsilon} (h_R(R) - \dot{\phi} - 3 \dot{R} h_{RR}(R)) - \epsilon \ddot{R} h_{RR}(R) - \epsilon \dot{R}^2 h_{RRR}(R)  \nonumber
\\
&&- \dot{\phi} f_{R \phi}(R,\phi)  +\ddot{\phi} f_{R \phi}(R,\phi) + 2 \dot{\phi}^2 f_{R \phi \phi}(R,\phi) - \dot{R} f_{RR}(R,\phi) + \ddot{R} f_{RR}(R,\phi)  \nonumber
\\
&&\left.+ 3 \dot{\phi} \dot{R} f_{RR \phi}(R,\phi) + \dot{R}^2 f_{RR}(R,\phi) \right]
= 0\,,
\label{EL R}
\end{eqnarray}
which can be solved after  the forms  of functions $h$ and $f$ are selected by the Noether symmetries. 


In this perspective, the generator of a generic transformation related to the minisuperspace defined by the configuration space ${\cal Q} \equiv \{a,\phi,R,\epsilon\}$  is
\begin{equation}
X = \xi(t,a,\phi,R,\epsilon) \partial_t + \alpha(t,a,\phi,R,\epsilon) \partial_a + \beta(t,a,\phi,R,\epsilon) \partial_\phi + \gamma(t,a,\phi,R,\epsilon) \partial_R + \delta(t,a,\phi,R,\epsilon) \partial_\epsilon\,.
\end{equation}
The identity \eqref{Teorema}, once applied to Lagrangian \eqref{R Lagr}, leads to a system of 19 equations, mostly connected by linear combinations. After neglecting redundant equations, it reduces to the below  system of  six differential equations: 

\begin{equation}
\begin{cases}
& 3 \alpha \left(f(R,\phi) - R f_R(R,\phi) - \epsilon h(R) + \epsilon R h_R(R) \right) + a \left[\delta(R h_R(R) - h(R)) + \beta f_\phi(R,\phi) - \beta R f_{R \phi}(R,\phi) \right. 
\\
&\left. - \gamma R f_{RR}(R,\phi) + \partial_t \xi \left(f(R,\phi) - R f_R(R,\phi) \right)  + \epsilon \gamma R h_{RR}(R)- \epsilon \partial_t \xi (h(R) - R h_R(R))\right] = 0\,;
\\
\\
&2 \alpha (f_{RR}(R,\phi) - \epsilon h_{RR}(R)) + a \left[- \delta h_{RR}(R) + \beta f_{RR\phi}(R,\phi) + \gamma f_{RRR}(R,\phi) - \partial_t \xi f_{RR}(R,\phi) + \partial_R \gamma f_{RR}(R,\phi)  \right.
\\
&\left. + \partial_a \alpha f_{RR}(R,\phi) - \epsilon (\gamma h_{RRR}(R) - \partial_t \xi h_{RR}(R) + \partial_R \gamma h_{RR}(R) + \partial_a \alpha h_{RR}(R)) \right] = 0\,;
\\
\\
& 12 \alpha f_{R\phi}(R,\phi) + 6a \left[\beta f_{R \phi \phi}(R,\phi) + \gamma f_{RR \phi}(R,\phi) - f_{R \phi}(R,\phi)(\partial_t \xi - \partial_\phi \beta - \partial_a \alpha) \right] + \partial_a \delta \;a^2  = 0\,;
\\
\\
& \alpha( \epsilon h_R(R) - f_R(R,\phi)) + a \left[h_R(R) \delta - \beta f_{R \phi}(R,\phi) - \gamma f_{RR}(R,\phi) + \partial_t \xi f_{R} (R,\phi) + \gamma \epsilon h_{RR}(R) - \partial_t \xi \epsilon h_R(R) \right.
\\
& + 2 \partial_a \alpha \epsilon h_R(R) - 2 \partial_a \alpha f_R(R,\phi)]- \partial_a \beta a^2 f_{R \phi} + \partial_a \delta \; a^2 h_R(R) = 0\,;
\\
\\
&12 \alpha h_R(R) + 6 a \left[\gamma h_{RR} + h_R \left(- \partial_t \xi + \partial_\epsilon \delta + \partial_a \alpha \right)\right] - \partial_a \beta a^2 = 0\,;
\\
\\
&3 \alpha - a \left(\partial_t \xi - \partial_\epsilon \delta - \partial_\phi \beta\right) = 0\,;
\\
\\
& \alpha = \alpha(a)\,, \;\;\; \beta = \beta(a,\phi)\,, \;\;\; \gamma = \gamma(R)\,, \;\;\; \delta = \delta(a,\epsilon)\,, \;\;\; \xi = \xi(t)\,, \;\;\; g=g_0\,.
\end{cases}
\end{equation}
The above system leads to three different classes of generators and functions:
\begin{equation}
\begin{cases}
I: \, & X = (-3\alpha_0 t + \xi_1) \partial_t + \alpha_0 a \partial_a  -6 \frac{\alpha_0}{n-1} R \partial_R + 6 \alpha_0 \epsilon \partial_\epsilon\,,
\\
& h(R) = h_0 R\,,  \;\;\;\;\; f(R,\phi)  =f_0R^n F(\phi)= f_0 R^n F(\Box^{-1} R)\,;
\\
\\
II: \, &X = \gamma_0 R \partial_R\,, 
\\
& h(R) = h_0 R\,,  \;\;\;\;\; f(R,\phi) = f_0 R + f_1 \phi^k = f_0 R^n + f_1 (\Box^{-1} R)^k \,;
\\
\\
III: \, &  X = (-3\alpha_0 t + \xi_1) \partial_t + \alpha_0 a \partial_a  -6 \frac{\alpha_0}{n-1} R \partial_R + \beta_0 \phi+ 6 \alpha_0 \epsilon \partial_\epsilon\,,
\\
&  h(R) = h_0 R\,,  \;\;\;\;\; f(R,\phi) = f_0 R^n e^{k \phi} = f_0 R^n e^{\Box^{-1} R}\,.
\end{cases}
\label{Solution Noether3}
\end{equation}
where only the last one admits analytic cosmological solutions. Replacing the form of the last functions into Eqs.~\eqref{EL R},  we  obtain the following de Sitter-like solution:
\begin{equation}
a(t) \sim e^{mt}\,, \,\,\,\,\, \phi(t) \sim t\,, \,\,\,\,\, R(t) \sim \text{Const.} \,\,\,\,\, \epsilon(t) \sim e^{-3 m t}\,,
\end{equation}
with the constraint
\begin{equation}
m = \frac{-3 + 10 h_0 k - 
 8 h_0^2 k^2}{2 (-3 + 18 h_0 k - 31 h_0^2 k^2 + 20 h_0^3 k^3)}\,.
\end{equation}
We can conclude that non-local curvature cosmology is consistent with an accelerated expansion of the scale factor of the universe. 

\section{Noether Symmetries in Non-local Gauss--Bonnet Cosmology} \label{Purenon}
A similar procedure can be applied to models containing only the GB invariant. As discussed in~\cite{Bajardi:2020osh}, a theory containing only combinations of $\G$ can reduce to GR, at least in a cosmological background.

Let us start by considering the non-local GB action in vacuum of the form
\begin{equation}
S= \int \sqrt{-g} f(\G, \Box^{-1} h(\G)) d^4x\,,
\label{initial action}
\end{equation}
from which we want to derive the cosmological point-like  Lagrangian. As above, we {\it localize} the theory by setting:
\begin{equation}
\Box^{-1} h(\G) := \phi \;\;\;\;\;\;\; \to \;\;\;\;\;\;\; h(\G) = \Box\phi \;.
\end{equation}
 By introducing a new scalar field $\epsilon(t)$ as in~\cite{Nojiri:2007uq, Elizalde:2018qbm}, the action can be written as
\begin{equation}
S= \int \sqrt{-g} \left\{f(\G, \phi) + \epsilon(t)(\Box \phi - h(\G)) \right\} d^4x\,,
\label{action in progress}
\end{equation}
where $\phi$ and $\G$ have to be  treated as separated fields. In a FLRW  spatially flat cosmology,  the GB invariant  turns out to be
\begin{equation}
\G = 24 \frac{\dot{a}^2 \ddot{a}}{a^3}.
\end{equation}
Therefore, by using the Lagrange Multipliers method,  action \eqref{action in progress} becomes
\begin{equation}
S= \int \left\{a^3 f(\G, \phi) + a^3 \epsilon(t)\left[\ddot{\phi} + 3 \frac{\dot{a}}{a} \dot{\phi} - h(\G)\right] - \lambda \left(\G - 24 \frac{\dot{a}^2 \ddot{a}}{a^3} \right) \right\}dt\,.
\end{equation}
The variation of the action with respect to the GB term allows to find the Lagrange Multiplier $\lambda$:
\begin{equation}
\frac{\delta S}{\delta \G} = \int \left\{ a^3 f_\G(\G,\phi) - a^3 \epsilon(t) h_\G(\G) - \lambda\right\} dt = 0 \;\;\;\;\; \to \;\;\;\;\;\; \lambda =  a^3 \left[f_\G(\G,\phi) - \epsilon(t) h_\G(\G) \right],
\end{equation}
where the subscript $\G$ denotes the derivative with respect to the GB scalar. Furthermore, from Eq.~\eqref{action in progress}, it is easy to verify that once varying the action with respect to $\epsilon(t)$, one recovers the definition $\Box \phi - h(\G) = 0$. From the variation with respect to the scalar field $\phi$, we get:
\begin{equation}
\frac{\delta S}{\delta \phi} = \int \sqrt{-g} \left\{ f_\phi(\G,\phi) - \frac{\delta}{\delta \phi} \left[\epsilon(t)(\Box \phi - h(\G))  \right]\right\} d^4 x = \int \sqrt{-g} \left\{f_\phi(\G,\phi) - \frac{\delta}{\delta \phi} \left[\epsilon(t)\Box (\phi - \Box^{-1} h(\G)) \right] \right\} d^4 x \;.
\label{variazphi}
\end{equation}
Using the divergence theorem, the last term of Eq.Eq.~\eqref{variazphi} can be written as:
\begin{eqnarray}
&& \int \sqrt{-g}\left\{\frac{\delta}{\delta \phi} \left[\epsilon(t)\Box (\phi - \Box^{-1} h(\G)) \right] \right\} d^4 x = \frac{\delta}{\delta \phi} \int \sqrt{-g} \Box \epsilon(t) (\phi - \Box^{-1} h(\G)) d^4x \nonumber 
\\
&& = \int \sqrt{-g} \; \Box \epsilon(t)  \frac{\delta}{\delta \phi} (\phi - \Box^{-1} h(\G)) d^4x = \int \sqrt{-g} \; \Box \epsilon(t) d^4x  \;,
\end{eqnarray}
so the variation with respect to the scalar field $\phi$ provides the following Klein-Gordon equation:
\begin{equation}
\frac{\delta S}{\delta \phi}  = 0 \;\;\;\;\;\; \to \;\;\;\;\;\; \Box \epsilon(t) = f_\phi(\G,\phi)\,.
\end{equation}
After introducing the Lagrange multipliers and integrating out the higher derivatives, the point-like Lagrangian can be written as:
\begin{equation}
\Lagr = a^3 \left[ f(\G,\phi) - \G f_\G(\G,\phi) - \epsilon h(\G) + \epsilon \G h_\G(\G) \right] - a^3 \dot{\phi} \dot{\epsilon} - 8 \dot{a}^3 \dot{\G} f_{\G\G}(\G,\phi) + 8 \dot{a}^3 \dot{\epsilon} h_\G(\G) + 8 \epsilon \dot{a}^3 \dot{\G} h_{\G\G}(\G) - 8 \dot{a}^3 \dot{\phi} f_{\G\phi}(\G,\phi)\,.
\label{GB lagr}
\end{equation}
The corresponding Euler--Lagrange equations and the energy condition are  respectively:
\begin{eqnarray}
\frac{d}{dt} \frac{\partial \Lagr}{\partial \dot{a}} = \frac{\partial \Lagr}{\partial a} \;\; \to \;\; &&  8 \dot{a} \left[2 \ddot{a} \left(-\dot{\G}f_{\G\G}(\G,\phi)-\dot{\phi}
  f_{\G \phi}(\G,\phi)+\epsilon \dot{\G} h_{\G\G}(\G)\right)+\dot{a} \left(-\ddot{\G}
  f_{\G\G}(\G,\phi) -2 \dot{\G} \dot{\phi} f_{\G\G \phi}(\G,\phi)  \right. \right. \nonumber
  \\
  && \left. -\dot{\G}^2
  f_{\G\G\G}(\G,\phi)-\ddot{\phi}f_{\G \phi}(\G,\phi)-\dot{\phi}^2
  f_{\G \phi \phi}(\G,\phi)+\ddot{\epsilon} h_\G(\G)+\epsilon \ddot{\G}
   h_{\G\G}(\G)+\epsilon \dot{\G}^2 h_{\G\G\G}(\G)\right)  \nonumber
   \\
   &&\left. +2 \dot{\epsilon} \left(\ddot{a}
   h_\G(\G)+\dot{a} \dot{\G} h_{\G\G}(\G)\right)\right]+a^2 \left[\G
  f_{\G}(\G,\phi)-f(\G,\phi)+\dot{\epsilon} \dot{\phi}+\epsilon
   \left(h(\G)-\G h_\G(\G)\right)\right]= 0\,; \nonumber
\end{eqnarray}
\begin{eqnarray}
\frac{d}{dt} \frac{\partial \Lagr}{\partial \dot{\phi}} = \frac{\partial \Lagr}{\partial \phi} \;\; \to \;\; && \Box \epsilon = f_\phi(\G, \phi)\,; \hspace{11cm} \nonumber
\end{eqnarray}
\begin{eqnarray}
\frac{d}{dt} \frac{\partial \Lagr}{\partial \dot{\G}} = \frac{\partial \Lagr}{\partial \G} \;\; \to \;\; && \G = 24 \frac{\dot{a}^2 \ddot{a}}{a^3}\,; \hspace{11.3cm} \nonumber
\end{eqnarray}
\begin{eqnarray}
\frac{d}{dt} \frac{\partial \Lagr}{\partial \dot{\epsilon}} = \frac{\partial \Lagr}{\partial_\epsilon} \;\; \to \;\; && \Box \phi = h(\G)\,; \hspace{11.3cm} \nonumber
\end{eqnarray}
\begin{eqnarray}
E_{\Lagr}=\dot{q}^i \frac{\partial \Lagr}{\partial \dot{q}^i} - \Lagr = 0 \;\; \to \;\; && -a^3 \left( f(\G,\phi) - \epsilon h(\G) + \epsilon \G h_\G(\G) + \dot{\phi} \dot{\epsilon} - \G f_\G(\G,\phi) \right)  \nonumber 
\\
&& +24 \dot{a}^3 \left( \dot{\epsilon} h_\G(\G) + \epsilon \dot{\G} h_{\G\G}(\G) - \dot{\phi} f_{\G \phi}(\G,\phi) - \dot{\G} f_{\G\G}(\G , \phi)\right) = 0\,.
\label{EL GB}
\end{eqnarray}
Once the forms of the functions $h(\G)$ and $f(\G,\phi)$ are specified, the above system can provide exact cosmological solutions.


Also here, the Noether theorem can be applied to the  Lagrangian \eqref{GB lagr}. In such a case the minisuperspace is defined on the configuration space ${\cal Q}\equiv\{a, \phi, \G, \epsilon\}$, so that the symmetry generator \eqref{generator} takes the explicit form:
\begin{equation}
X = \xi(t,a,\phi,\G,\epsilon) \partial_t + \alpha(t,a,\phi,\G,\epsilon) \partial_a + \beta(t,a,\phi,\G,\epsilon) \partial_\phi + \gamma(t,a,\phi,\G,\epsilon) \partial_\G + \delta(t,a,\phi,\G,\epsilon) \partial_\epsilon\,.
\end{equation} 
The system of differential equations coming from the above generator is made of 37 equations but, after deleting all the linear combinations, it reduces to a system of five equations plus the conditions on the generator coefficients:
\begin{equation}
\begin{cases}
&\displaystyle 3 \alpha a^2 f(\G,\phi) - \delta a^3 - 3 \alpha a^2 \epsilon h(\G) + \delta a^3 \G h_\G(\G)  + 3 \alpha a^2 \epsilon \G h_\G(\G) + \gamma a^3 \G \epsilon h_{\G\G}(\G) + \beta a^3 f_\phi(\G,\phi) - 3 \alpha a^2 \G f_\G(\G,\phi) 
\\
&- \beta a^3 \G f_{\G \phi}(\G,\phi) - \gamma a^3 \G f_{\G\G}(\G,\phi) - \partial_t g + a^3 \partial_t \xi \left( f(\G,\phi) - \epsilon h(\G) + \epsilon \G h_\G(\G) - \G f_\G(\G,\phi) \right) = 0\,;
\\
\\
&\displaystyle \gamma h_{\G\G}(\G) - 3 \partial_t \xi h_\G(\G) + \partial_\epsilon \delta \; h_\G(\G) + \partial_\epsilon \gamma \left(\epsilon h_{\G\G}(\G) - f_{\G\G}(\G,\phi) \right) + 3 \partial_a \alpha h_\G(\G) = 0\,;
\\
\\
&\displaystyle f_{\G \phi}(\G,\phi) \left(3 \partial_t \xi - \partial_\phi \beta \right) + \partial_\phi \gamma \left(\epsilon h_{\G\G}(\G) - f_{\G\G}(\G,\phi) \right) - 3 \partial_a \alpha f_{\G\phi}(\G,\phi) - \beta f_{\G \phi \phi} - \gamma f_{\G \G \phi}(\G, \phi) = 0\,;
\\
\\
&\displaystyle \beta f_{\G \G \phi}(\G,\phi) - \delta h_{\G \G}(\G) + \gamma f_{\G\G\G}(\G,\phi) - 3 \partial_t \xi f_{\G \G}(\G,\phi) + \partial_\G \gamma f_{\G\G}(\G,\phi) + 3 \partial_a \alpha f_{\G \G}(\G,\phi) - \gamma \epsilon h_{\G\G\G}(\G) + 3 \partial_t \xi \epsilon h_{\G \G} 
\\
& - \partial_\G \gamma \epsilon h_{\G\G}(\G) - 3 \partial_a \alpha \epsilon h_{\G\G}(\G) = 0\,;
\\
\\
&\displaystyle 3 \alpha - a \left(\partial_t \xi - \partial_\epsilon \delta - \partial_\phi \beta \right) = 0\,;
\\
\\
&\displaystyle \alpha \equiv \alpha(a)\,, \;\;\;\; \beta \equiv \beta(\phi)\,, \;\;\;\; \gamma \equiv \gamma(\phi,\G,\epsilon)\,, \;\;\;\; \delta \equiv \delta(\epsilon)\,, \;\;\;\; \xi \equiv \xi(t)\,, \;\;\;\; g \equiv g(t)\,.
\end{cases}
\end{equation}
The above system admits five different solutions: in two of them the non-local function $f(\G,\phi)$ is given by a sum of a function of $\phi$ and a function of $\G$, \emph{i.e.} $f(\G,\phi) = f_1(\G) + f_2 (\phi)$. In the other three,  solutions are  products between the two functions, namely $f(\G,\phi) = g_1(\G) g_2(\phi)$. The entire set of solutions with the corresponding generators read:
\begin{equation}
\begin{cases}
I: \, &X = (\xi_0 t + \xi_1) \partial_t + \alpha_0 a \partial_a + (\beta_0 \phi + \beta_1) \partial_\phi -4\xi_0 \G \partial_\G + \delta_0 \epsilon \partial_\epsilon\,,
\\
& h(\G) = h_0 \G^{\frac{1}{2} + \frac{n}{k}}\,, \;\;\;\;\; f(\G,\phi) = f_0 \G^n + f_1 \G + f_2 \left(\beta_0  \phi + \beta_1\right)^k = f_0 \G^n + f_1 \G + f_2 \left(\beta_0 + \beta_1 \Box^{-1} \G^{\frac{1}{2} + \frac{n}{k}} \right)^k\,;
\\
\\
II: \, &X = (\xi_0 t + \xi_1) \partial_t + \alpha_0 a \partial_a + (\beta_0 \phi + \beta_1) \partial_\phi -4\xi_0 \G \partial_\G + (\delta_0 \epsilon + \delta_1) \partial_\epsilon\,,
\\
& h(\G) = h_0 \G\,,  \;\;\;\;\; f(\G,\phi) = f_0 \G^n + f_1 \G + f_2 (\beta_0 \phi + \beta_1)^{2n} = f_0 \G^n + f_1 \G + f_2 (\beta_0 \Box^{-1} \G + \beta_1)^{2n}\,;
\\
\\
III: \, &X = (\xi_0 t + \xi_1) \partial_t + \alpha_0 a \partial_a + (\beta_0 \phi + \beta_1) \partial_\phi - 4\xi_0 \G \partial_\G + \delta_0 \epsilon \partial_\epsilon\,, 
\\
& h(\G)=h_0 \G^z\,, \;\;\;\;\; f(\G,\phi) = f_0 \G^n  (\beta_0 \phi + \beta_1)^k =  f_0 \G^n  (\beta_0 \Box^{-1} \G^z + \beta_1)^k\,;
\\
\\
IV: \, &X = (\xi_0 t + \xi_1) \partial_t + \alpha_0 a \partial_a + (\beta_0 \phi + \beta_1) \partial_\phi -4\xi_0 \G \partial_\G + (\delta_0 \epsilon + \delta_1) \partial_\epsilon\,,
\\
& h(\G) = h_0 \G\,, \;\;\;\;\; f(\G,\phi) = f_0 \G^n (\beta_0 \phi + \beta_1)^k = f_0 \G^n (\beta_0 \Box^{-1} \G + \beta_1)^k\,;   
\\
\\
V: \, &X = (\xi_0 t + \xi_1) \partial_t + \alpha_0 a \partial_a  +\beta_1 \partial_\phi - 4 \xi_0 \G \partial_\G + \delta_0 \epsilon \partial_\epsilon\,,
\\
&h(\G) = h_0 \sqrt{\G}\,, \;\;\;\;\; f(\G,\phi) = f_0 \G^n e^{k \phi} = f_0 \G^n e^{k \Box^{-1} \sqrt{\G}}\,,    \; \,\,\,\,\, k \equiv \frac{\delta_0 + 4 n \xi_0}{\beta_1}\,,
\end{cases}
\label{Solution Noether1}
\end{equation}
where $\xi_0, \, \xi_1, \, \alpha_0, \, \beta_0, \, \beta_1, \, \delta_0, \, h_0, \, f_0, \, f_1, \, f_2, \, n, \, k$ are integration constants.

It may seem that the theory is over determined by the large amount of free parameters. However, after solving the equations of motion, the functions will be further constrained to those in agreement with the cosmological solutions. Specifically, it turns out that not all the functions contained in the system \eqref{Solution Noether1} admit cosmological solutions for the scale factor. As a matter of fact, while the second and the fourth do not admit any cosmological solution, the first and the third  can be analytically solved by setting $\beta_1 = 0$. The fifth admits solutions only after constraining the mutual dependence among the parameters. 

Let us start by analyzing the Lagrangians corresponding to the functions I and III. They read, respectively:
\begin{eqnarray}
\Lagr_I &=& a^3 \left[f_0 (1-n) \G^n - h_0\left( \frac{n}{k}-\frac{1}{2}\right) \epsilon \G^{\frac{n}{k} + \frac{1}{2}} +\dot{\epsilon} \dot{\phi} \right] - 8 h_0 \left(\frac{n}{k} + \frac{1}{2} \right) \dot{a}^3 \dot{\epsilon} \G^{\frac{n}{k} - \frac{1}{2}} \nonumber
\\
&&+ 8 f_0 n(n-1) \dot{a}^3 \dot{\G} \dot{\G}^{n-2} - 8 h_0 \left(\frac{n^2}{k^2} - \frac{1}{4} \right) \epsilon \dot{a}^3 \dot{\G} \G^{\frac{n}{k} - \frac{3}{2}} +8f_1 k \dot{a}^3 \dot{\phi} \phi^{k-1}\,,  
\label{sum Lagr}
\end{eqnarray}
and
\begin{eqnarray}
\Lagr_{III} &=& a^3 \left[f_0 (1-n) \G^n \phi^k  + h_0 (z-1) \epsilon \G^z \right] + 8 h_0 z \dot{a}^3 \dot{\epsilon}  \G^{z-1} +  8 h_0 z (z-1) \epsilon \dot{a}^3 \dot{\G}  \G^{z-2}  \nonumber
\\
&& - 8 f_0 n(n-1) \dot{a}^3 \dot{\G} \G^{n-2} \phi^k - 8 f_0 k n \dot{a}^3 \dot{\phi} \G^{n-1} \phi^{k-1} - a^3 \dot{\epsilon} \dot{\phi}\,.
\label{prod lagr}
\end{eqnarray}
From the former Lagrangian, the Euler--Lagrange equations and the energy condition provide a solution given by a time power-law of $a(t)$, namely:
 \begin{eqnarray}
  && a(t) \sim t^{\frac{2}{3}(2n+2kz -k)}\,, \;\;\; \G(t) \sim t^{-4}\,, \;\;\; \phi(t) \sim  t^{2-4z}\,, \;\;\; \epsilon(t) \sim t^{2k(1-2z)}\,, 
  \\
  && \;\;\;\;\;\;\;\;\;\;\;\;\; f(\G, \Box^{-1} h(\G)) = f_2  \G^n (\Box^{-1} \G^z)^k\,.
 \end{eqnarray}
In this case, though exponential solutions do not occur in vacuum, the parameters are not fixed by the equations of motion, so that they might be constrained by observations. On the other hand, the Lagrangian \eqref{sum Lagr} gives exact de Sitter-like solutions of the form:
\begin{eqnarray}
&& a(t) \sim e^{q t}\,, \;\;\; \G(t) \sim \text{Const}\,, \;\;\; \phi(t) \sim t\,, \;\;\; \epsilon(t) \sim t\,, \;\;\; k=1, \; n= \frac{1}{2}\,, 
\\
&& \;\;\;\;\;\;\;\;\;\;\;\;\; f(\G, \Box^{-1} h(\G)) = f_0 \sqrt{\G} + f_1 \G + f_2  \Box^{-1} \G + f_3\,.
\end{eqnarray}

Therefore, concerning this latter case, the only solution such that the Euler--Lagrange equations, the energy condition, and the Noether system are satisfied, constrains all the free parameters occurring in the second function. Since $\G$ is a topological invariant,  the linear term in $\G$ does not contribute to  dynamics  so the relevant terms are  the square root  and the  linear non-local terms in $\G$. From a cosmological point of view, this action is equivalent to  action~\eqref{NL:8}, so the same considerations in~\cite{Deser:2007jk} hold. In other words, a dark energy-like behavior, due to non-local terms, can be achieved both in  $R$ and $\G$ descriptions of cosmological dynamics. Furthermore, as discussed in~\cite{Bajardi:2020xfj}, different theories exhibiting the same Noether symmetries have the same dynamics.

Finally, the  point-like Lagrangian corresponding to the last solution  is:
\begin{eqnarray}
\Lagr_{V} &=& 2 \dot{a}^3 \left[2 \G^{-\frac{1}{2}} \dot{\epsilon} -\epsilon \G^{-\frac{3}{2}} \dot{\G} -4 f_0 n (n-1)  \G^{n-2} \dot{\G} e^{k \phi}-4 f_0 k n \G^{n-1} e^{k \phi} \dot{\phi} \right] 
\nonumber
 \\
 && -\frac{1}{2} \G^2 a^3 \left[2 \dot{\epsilon} \dot{\phi} + \epsilon \sqrt{\G}+2 f_0 (n-1) \G^n e^{k \phi}\right],
 \label{Lagr5} 
\end{eqnarray}
and the Euler--Lagrange Eqs.~\eqref{EL GB} can be analytically solved providing two different forms of the scale factor; the first reads as:
\begin{eqnarray}
&&\displaystyle a(t) \sim e^{q \, t}\,, \,\,\,\,\ \phi(t)\sim  t\,, \,\,\,\,\,\, \epsilon(t) \sim  e^{\sqrt{\frac{8}{3}}k q \, t}\,, 
\\
&& \displaystyle f(\G,\Box^{-1} \sqrt{\G}) = f_0 \G^{\frac{12 \sqrt{6}}{4k - \sqrt{6}}} e^{k \phi}\,, \,\,\,\,\,\,\,\,\,\, h(\G) =\sqrt{\G}\,.
\label{func5}
\end{eqnarray}
By comparing the function in Eq.~\eqref{func5} with the fifth of Eq.~\eqref{Solution Noether1}, we notice that a relation between the free parameters $n$ and $k$ occurs, namely:
\begin{equation}
n = \frac{12 \sqrt{6}}{4k - \sqrt{6}}\,.
\end{equation}
Furthermore, we also find  power law solutions, namely:
\begin{equation}
a(t) \sim  t^q\,, \,\,\,\,\,\, \phi(t) \sim  \ln[(1 - 3 q ) t]\,, \,\,\,\,\,\,\,\, \G(t) \sim \frac{1}{t^4}\,, \,\,\,\,\,\,\, \epsilon(t) \sim t^{2-4n + \frac{2 k \sqrt{6 q^3 (q-1)}}{3q-1}}\,.
\end{equation}
The second solution, coming from the Lagrangian in Eq.~\eqref{Lagr5},  introduces a relation among the parameters $n$, $q$ and $k$ enlarging  the possibility to compare these cosmological behaviors with observational data. 

\subsection{Noether Symmetries in General Relativity plus Non-local Gauss--Bonnet Cosmology}\label{Noether Symmetry Approach3}

To conclude this discussion, let us treat the case of GR corrected with non-local GB terms, considered \emph{e.g.} in~\cite{Elizalde:2018qbm, Capozziello:2008gu}.
The  action  is:
\begin{equation}
S = \int \sqrt{-g} \left[\frac{R}{2 \kappa} + f(\G, \Box^{-1} h(\G) \right] d^4x\,,
\label{action2}
\end{equation}
which is nothing but action  \eqref{initial action} with the addition of the Einstein--Hilbert term. Let us make use of the Lagrange multipliers method to find the cosmological Lagrangian and, therefore, to apply the Noether Symmetry approach. The only difference with respect to the case given by Eq.~\eqref{GB lagr} is due to the cosmological form of $R$, so that the Lagrangian reads as:
\begin{eqnarray}
\Lagr &=& a^3 \left[ f(\G,\phi) - \G f_\G(\G,\phi) - \epsilon h(\G) + \epsilon \G h_\G(\G) \right] - a^3 \dot{\phi} \dot{\epsilon} - 8 \dot{a}^3 \dot{\G} f_{\G\G}(\G,\phi)  \nonumber
\\
&&+ 8 \dot{a}^3 \dot{\epsilon} h_\G(\G) + 8 \epsilon \dot{a}^3 \dot{\G} h_{\G\G}(\G) - 8 \dot{a}^3 \dot{\phi} f_{\G\phi}(\G,\phi) + \frac{3}{\kappa} a \dot{a}^2\,,
\end{eqnarray}
and the only different Euler--Lagrange equation is that related to the scale factor which, in this case, takes the form:
\begin{eqnarray}
\frac{d}{dt} \frac{\partial \Lagr}{\partial \dot{a}} = \frac{\partial \Lagr}{\partial a} \;\; \to \;\; &&  8 \dot{a} \left[2 \ddot{a} \left(-\dot{\G}f_{\G\G}(\G,\phi)-\dot{\phi}
  f_{\G \phi}(\G,\phi)+\epsilon \dot{\G} h_{\G\G}(\G)\right)+\dot{a} \left(-\ddot{\G}
  f_{\G\G}(\G,\phi) -2 \dot{\G} \dot{\phi} f_{\G\G \phi}(\G,\phi)  \right. \right. \nonumber
  \\
  && \left. -\dot{\G}^2
  f_{\G\G\G}(\G,\phi)-\ddot{\phi}f_{\G \phi}(\G,\phi)-\dot{\phi}^2
  f_{\G \phi \phi}(\G,\phi)+\ddot{\epsilon} h_\G(\G)+\epsilon \ddot{\G}
   h_{\G\G}(\G)+\epsilon \dot{\G}^2 h_{\G\G\G}(\G)\right)  \nonumber
   \\
   &&\left. +2 \dot{\epsilon} \left(\ddot{a}
   h_\G(\G)+\dot{a} \dot{\G} h_{\G\G}(\G)\right)\right]+a^2 \left[\G
  f_{\G}(\G,\phi)-f(\G,\phi)+\dot{\epsilon} \dot{\phi}+\epsilon
   \left(h(\G)-\G h_\G(\G)\right)\right]  \nonumber
   \\
   && + \frac{3}{\kappa} (\dot{a}^2 + 2 a \ddot{a})= 0 \,.
\end{eqnarray}
The minisuperspace dimension is the same as the previous case, since the scalar curvature does not introduce any new dynamical variables. By replacing $R(t)$ with its cosmological expression, the Noether system turns out to be the same of that of the previous section, except for the addition of a further condition on the Noether vector, that is:
\begin{equation}
\alpha - a \partial_t \xi + 2 a \partial_a \alpha=0\,.
\label{furtheq}
\end{equation}
This new link between $\alpha$ and $\xi$, provided by Eq.~\eqref{furtheq}, yields an important implication for the solutions of Noether's system, since it uniquely fixes the value of $n$ and the relation between $n$ and $k$. We obtain five different generators (with corresponding functions) of the form:
\begin{equation}
\begin{cases}
I: \, & X = (3\alpha_0 t + \xi_1) \partial_t + \alpha_0 a \partial_a + (\beta_0 \phi + \beta_1) \partial_\phi -12\alpha_0 \G \partial_\G + \delta_0 \epsilon \partial_\epsilon\,,
\\
& h(\G) = h_0 \G^{\frac{1}{2} + \frac{1}{2k}}\,, \;\;\;\;\; f(\G,\phi) = f_0 \G^\frac{1}{2} + f_1 \G + f_2 \left(\beta_0  \phi + \beta_1\right)^k = f_0 \G^\frac{1}{2} + f_1 \G + f_2 \left(\beta_0 + \beta_1 \Box^{-1} \G^{\frac{1}{2} + \frac{1}{2k}} \right)^k\,;
\\
\\
II: \, & X = (3\alpha_0 t + \xi_1) \partial_t + \alpha_0 a \partial_a + (\beta_0 \phi + \beta_1) \partial_\phi -12\alpha_0 \G \partial_\G + (\delta_0 \epsilon + \delta_1) \partial_\epsilon\,,
\\
& h(\G) = h_0 \G\,,  \;\;\;\;\; f(\G,\phi) = f_0 \G^\frac{1}{2} + f_1 \G + f_2 (\beta_0 \phi + \beta_1) = f_0 \G^\frac{1}{2} + f_1 \G + f_2 (\beta_0 \Box^{-1} \G + \beta_1)\,;
\\
\\
III: \, & X = (3\alpha_0 t + \xi_1) \partial_t + \alpha_0 a \partial_a + (\beta_0 \phi + \beta_1) \partial_\phi - 12\alpha_0\G \partial_\G + \delta_0 \epsilon \partial_\epsilon\,, 
\\
& h(\G) = h_0 \G^{\frac{1-2n}{2k}}\,, \;\;\;\;\; f(\G,\phi) = f_0 \G^n  (\beta_0 \phi + \beta_1)^k =  f_0 \G^n  (\beta_0 \Box^{-1} \G^{\frac{1-2n}{2k}} + \beta_1)^k\,;
\\
\\
IV: \, & X = (3\alpha_0 t + \xi_1) \partial_t + \alpha_0 a \partial_a + (\beta_0 \phi + \beta_1) \partial_\phi - 12\alpha_0 \G \partial_\G + (\delta_0 \epsilon + \delta_1) \partial_\epsilon\,,
\\
& h(\G) = h_0 \G\,, \;\;\;\;\; f(\G,\phi) = f_0 \G^n (\beta_0 \phi + \beta_1)^{1-2n} = f_0 \G^n (\beta_0 \Box^{-1} \G + \beta_1)^{1-2n}\,;  
\\
\\
V: \,& X = (3\alpha_0 t + \xi_1) \partial_t + \alpha_0 a \partial_a + \beta_1\partial_\phi - 12\alpha_0 \G \partial_\G\,, 
\\
& h(\G) = h_0 \sqrt{\G}\,, \;\;\;\;\; f(\G,\phi) = f_0 \G^n e^{k \phi} = f_0 \G^n e^{k \Box^{-1} \sqrt{\G}}\,.
\end{cases}
\label{Solution Noether2}
\end{equation}
By comparing Eqs.~\eqref{Solution Noether1} with Eqs.~\eqref{Solution Noether2}, we notice that, with regards to the first two solutions, the introduction of $R$ leads to the further constraint $n=1/2$. Moreover, in the third and in the fourth cases, the further relations $ z=\frac{1-2n}{2k}$ and $k=1-2n$ occur respectively. According to these considerations, it is clear that action \eqref{action2} is fully consistent with \eqref{initial action} and then it is not necessary introduce by hand  the Einstein-Hilbert term to recover GR in this context.

\section{Discussion and Conclusions} \label{Conclusions}
The existence of Noether symmetries is a powerful tool to reduce and solve dynamics in a wide class of physical problems. In particular, the method proves  very useful in cosmology to select the form of effective  Lagrangians and to solve the related dynamical systems. Here, we considered non-local theories of gravity where the curvature scalar $R$ and the topological invariant $\G$ are involved. Noether symmetries allow to fix the form of cosmological  Lagrangians, to reduce the dynamics  and to find exact solutions. It is important to stress that non-local terms, selected by symmetries, are of physical interest and allow renormalization and unitarity in higher derivative and non-local gravity theories~\cite{Modesto:2017hzl,Tomboulis:2015esa}. According to this statement, the existence of Noether symmetries can be considered a  physical criterion to select physical models.

Furthermore, many of cosmological solutions found are in agreement with the previous statement by Deser and Woodard~\cite{Deser:2007jk} that non-local cosmology can reproduce dark energy behavior at IR scales. Among the classes of considered models, the  action $S = \int \sqrt{-g} f[\G,\Box^{-1} h(\G)] d^4x$ presents an interesting phenomenology because generalizes the analogue  $S = \int \sqrt{-g} f[R,\Box^{-1} h(R)] d^4x$ and admits also the possibility to recover GR corrected with non-local terms. In this perspective, considering gravitational actions involving  the topological invariant $\G$  seems  extremely useful to cure and fix problems that arise from taking into account other curvature invariants.

In a forthcoming paper, we will compare the above solutions with observations in view of selecting reliable models to reconstruct a self-consistent cosmic history.

\section*{Acknowledgements}
The Authors  are supported  by the INFN sezione di Napoli, {\it iniziative specifiche} GINGER, MOONLIGHT2, QGSKY and TEONGRAV. 

\appendix

\begin{appendix}

\section{The Noether Symmetry Approach}
 \label{Noether Symmetry Approach}

Let us shortly recall the main properties of the Noether Symmetry Approach showing the method used to find  symmetries in the above theories of gravity. We adopted it to select the functional form of the  gravitational actions, to reduce dynamics and, finally, to find out exact solutions.

A general coordinate transformation with generator
\begin{equation}
X =  \xi \frac{\partial }{\partial t} + \eta^i \frac{\partial }{\partial q^i} \;,
\label{generator}
\end{equation} 
 preserving the Euler--Lagrange equations, must satisfy the relation:
\begin{equation}
X^{[1]} \Lagr + \dot{\xi} \Lagr = \dot{g}(t,q^i) \;,
\label{Teorema}
\end{equation}
with $X^{[1]}$ being defined as:
\begin{equation}
X^{[1]} =  \xi \frac{\partial }{\partial t} + \eta^i \frac{\partial }{\partial q^i} + (\dot{\eta}^i - \dot{q}^i \dot{\xi}) \frac{\partial}{\partial \dot{q}^i} \;,
\end{equation} 
where $g$ is a generic function of the variables of the considered minisuperspace, that is the space of configurations ${\cal Q}\equiv\{q_i\}$ whose tangent space is $T{\cal Q}\equiv\{q^i,\dot{q}^i\}$. Furthermore, if the condition in Eq.~\eqref{Teorema} holds,  a conserved quantity, depending on the variables and on their first derivatives, can be found; specifically, it can be shown that it is:
\begin{equation}
I(t,q^i,\dot{q}^i) = 	\displaystyle \xi \left(\dot{q}^i \frac{\partial \Lagr}{\partial \dot{q}^i} - \Lagr \right) - \eta^i \frac{\partial \Lagr}{\partial \dot{q}^i} + g(t,q^i) \;,
\end{equation}
where $I(t,q^i,\dot{q}^i)$ is a first integral of motion i.e. a conserved quantity. If the conserved quantity exists, it is possible to find out a coordinate system where a cyclic variable appears. The procedure allows to reduce dynamics and, eventually, to solve it finding out exact solutions as in the cases discussed above.
\end{appendix}


\end{document}